\documentclass[aps,prd,preprint,superscriptaddress,nofootinbib]{revtex4-1}
\pdfoutput=1
\usepackage{graphicx}  % needed for figures
\usepackage{latexsym}
\usepackage{slashed}
\usepackage{dcolumn}   % needed for some tables
\usepackage{bm}        % for math
\usepackage{amsmath}
\usepackage{amssymb}   % for math
\usepackage{longtable}
\usepackage{float}

\usepackage{mathtools}
\DeclarePairedDelimiter\bra{\langle}{\rvert}
\DeclarePairedDelimiter\ket{\lvert}{\rangle}
\DeclarePairedDelimiterX\braket[2]{\langle}{\rangle}{#1 \delimsize\vert #2}

\newcommand{\lc}{\affiliation{Department of Chemistry and Physics, LaGrange College, LaGrange, GA 30240, USA}}

\begin{document}

%\title{Continuity Equation Coupling $Q$ Derived from the Boltzmann Equation for Coupled Dark Matter and Dark Energy}
\title{Deriving the Dark Matter-Dark Energy Interaction Term in the Continuity Equation from the Boltzmann Equation}
%\vskip .45in

\author{Kevin J. Ludwick}
\email{kludwick@lagrange.edu} \lc
\author{Holston Sebaugh} 
\email{hsebaugh@lagrange.edu} \lc

\begin{abstract}

Dark energy and dark matter are two of the biggest mysteries of modern cosmology, and our understanding of their fundamental nature is incomplete.  Many parameterizations of couplings between the two in the continuity equation have been studied in the literature, and observational data from the growth of perturbations can constrain these parameterizations.  
%Dark matter and dark energy interact gravitationally %over long distances, so they should at least be coupled via the graviton.  
Assuming standard general relativity with a simple Yukawa-type coupling between dark energy and dark matter fields in the Lagrangian, 
we use the Boltzmann equation to analytically express and calculate the interaction kernel $Q$ in the continuity equation and compare it to that of a typical parametrization.  We arrive at a comparably 
very small result, as expected.  %For a reasonable range of dark matter mass, $1$ GeV $\leq M \leq 1$ TeV, $Q$ ranges from $10^{-100}$ eV$^5$ to $10^{-80}$ eV$^5$ as constrained by cosmological data.
%$1$ MeV $\leq M \leq 1$ TeV, $Q$ ranges from $10^{-109}$ eV$^5$ to $10^{-67}$ eV$^5$ as constrained by cosmological data.  
Since the interaction is a function of the dark matter mass, other observational data sets can be used to constrain the mass.  This calculation can be modified to account for other couplings of the dark energy and dark matter fields.     
%examine any observational implications that may follow.  
This calculation required obtaining a distribution function for dark energy that leads to an equation of state parameter that is negative, which neither Bose-Einstein nor Fermi-Dirac statistics can supply, and this is the main result of this paper.  Treating dark energy as a quantum scalar field, we use adiabatic subtraction to obtain a finite analytic approximation for its distribution function that assumes the FLRW metric and nothing more.  

%*Could use scattering amplitude in curved spacetime (from Birrel and Davies textbook) or try quantum Boltzmann equation. 

\end{abstract}

\maketitle

\renewcommand{\thepage}{\arabic{page}}
\setcounter{page}{1}
%\renewcommand{\thefootnote}{\#\arabic{}}

%\newpage

\begin{center}
{\bf Introduction}
\end{center}

It is now well known that some form of dark energy, comprising about 70\% of our universe, is responsible for cosmic acceleration, and that dark matter is the next most prevalent non-luminous substance in our universe, comprising about 25\% of it.  The fundamental nature of dark matter and dark energy and how they interact with each and the Standard Model is uncertain.  However, interaction between 
dark matter and dark energy can be constrained by the matter power spectrum \cite{1809.00550, 1809.02411, 1812.05493, 1812.06854, 1812.05594} .  Usually, an {\it ad hoc} parametrization for an interaction between dark energy (DE) and dark matter (DM)
as perfect fluids is assumed since we don't know of any fundamental coupling between them.  Conservation of energy-momentum implies the continuity equation:
\begin{equation}
\label{Econs}
\nabla_\alpha  T^{\mu \nu} = 0 ~~~ \rightarrow  ~~~~ \Sigma_i (\dot{\rho_i} + 3H (\rho_i+p_i) )= 0,
\end{equation}
where $\rho$ is the energy density, $p$ is the pressure, and $H$ is the Hubble parameter, and the sum is over the DE and DM components of the universe for 
late cosmological times since they dominate.  
If there is an interaction between DE and DM, we would have
\begin{equation}
\label{interactionQ}
\dot{\rho_{DM}} + 3H (\rho_{DM}+p_{DM}) = Q, ~~~ \dot{\rho_{DE}} + 3H (\rho_{DE}+p_{DE}) = -Q.
\end{equation}

Interaction between DE and DM can ameliorate the coincidence problem and the $H_0$ tension, and different parameterizations for the interaction kernel $Q$ have been widely studied in the literature \cite{1908.03663, 9908023, 1812.06854, 1310.0085, 1603.08299, 1706.04953, 1809.06883, 1805.08252, 1905.08286, 1908.04281, 1501.06540, 1502.04030, 1501.03073, 1807.05541, 1905.10382, 0606520, 1507.00187, 0609597, 0610806, 0702015, 0801.4233, 9408025}.  Influences for the parametrizations in the literature include computational convenience, ease of data comparison, and analogies from field theory.  The uncertain nature of the field theory of dark energy helps motivate the phenomenological approach to interaction kernels \cite{1110.3045, 0706.3064}.  
%mention data preference over non-interaction; attractor parametrizations
%including a variational formulation for relativistic fluids 
%\cite{1501.06540, 1502.04030}, 
%disformal couplings 
%\cite{1501.03073, 1807.05541}, , 1905.10382}.  
These parametrizations are usually convenient choices that allow for analytic solutions of Friedmann's equations.   In this work, we are interested in what $Q$ would look like if a coupling on a more fundamental 
level were assumed, namely, a coupling as fields in the Lagrangian.  Field couplings between DE and DM have been studied in the literature \cite{9908023, 1603.08299, 2006.04618, 1605.00996, 1910.02699, 1808.05015, 1411.3660}, 
%1605.00996: axion for DE and some of DM allows for non-negligible heavy DM coupled to DE; 1910.02699: non-minimal scalar field DE and kinetic coupling, Horndeski
but what we are interested in is a way of expressing an interaction in terms of a distribution function that only takes effect at 1st order and above in cosmological perturbation theory, and this scheme is typically carried out by expressing the energy density and pressure of fluid components in terms of a weighted integral of the distribution function \cite{9506072}.  Field coupling interactions are then taken into account via the collision term in the Boltzmann equation.  

Typically, dynamical DE is modeled as a scalar field, and the usual way of ensuring a slowly varying field that provides sufficient late-time cosmic acceleration in accordance with observations is to let the mass be on the order of $H_0$ or less.  In this case, if DE were in thermodynamic equilibrium with the rest of the contents of the early universe, it would be a Bose-Einstein condensate \cite{1411.0753}, and expressing the energy density and pressure as weighted integrals of a distribution function would not be valid since the majority of the energy would be in the ground state; a discrete sum instead of a continuous integral would be needed.  However, we assume here that DE is not a condensate, and our methodology that follows serves more as a proof of concept for DE and other general scalar field theories rather than a thorough investigation of the nature of DE.  

We let the scalar field $\phi$ represent DE and the scalar field $\psi$ represent DM.  Assuming a Yukawa-type coupling (last term) in the action 
for late times in which DE and DM dominate, the action is
\begin{equation}
\label{action}
S = \int d^4 x \sqrt{-g} \left[ \frac{R}{16 \pi G} - \frac{1}{2} g^{\mu \nu} \nabla_\mu \phi \nabla_\nu \phi - V(\phi) - \frac{1}{2} \xi R \phi^2 - \frac{1}{2} g^{\mu \nu} \nabla_\mu \psi \nabla_\nu \psi - V(\psi) - \frac{1}{2} \chi R \psi^2  - \frac{1}{2} g \psi^2 \phi \right] ,
\end{equation}
in which $\xi$ and $\chi$ are the non-minimal coupling parameters for DE and DM respectively, and they are in general present formally due to renormalization of the scalar fields.  The Yukawa-type coupling parameter $g$ is assumed to be small and perturbative.  

In what follows, instead of assuming a certain parametrization for $Q$, we calculate what $Q$ should be for this Yukawa-type coupling by utilizing the tree-level 
cross section with the Boltzmann equation.  In theory, a weakly interacting scalar field would have a pressure given by a weighted sum or integral involving the Bose-Einstein statistics factor, which would result in a positive pressure.  However, DE pressure must be negative when modeled as a perfect fluid according to the cosmic acceleration requirements.  It is also known that the correspondence between a scalar field theory and a perfect fluid representation is not perfect \cite{1201.1448}.  In order to deal with this incompatibility, we must find a suitable effective distribution function for DE that results in a negative pressure. 
%In the non-relativistic limit, both lead to an equation of state parameter $w \approx 0$ so that $p \approx 0$, as is the case for dark and luminous matter.  
We use adiabatic expansion in the next section to arrive at an approximate expression for the distribution function for DE, and we calculate $Q$ in the section after that.  
Then we analyze our result and compare it to observational constraints on a typical parameterization in the literature, and then we conclude.  

In this work, we assume the FLRW metric $ds^2 = -dt^2 +a(t)^2 (dx^2+dy^2+dz^2)$, and we assume $\hbar = c = k_B = 1$.  

\bigskip
\begin{center}
{\bf Distribution Function for Dark Energy}
\end{center}

In this section, we calculate an effective distribution function for dark energy that is compatible with negative pressure.  We do this by quantizing the scalar field theory for dark energy and finding the renormalized expectation value for the energy density obtained from the stress-energy tensor.  We regularize and renormalize using adiabatic subtraction.  We then compare our result with the expression for energy density from statistical mechanics to identify the effective distribution function that we will later use in the Boltzmann equation.

The dark energy fluid must have a negative pressure, but DE as a scalar field would have the Bose-Einstein distribution function, which would result in positive pressure according to the statistical mechanics expression for pressure, 
\begin{equation}
\label{pressure}
p = \frac{1}{a^4} \int \frac{d^3k}{(2 \pi)^3} \frac{k^2}{3 \sqrt{k^2+ a^2 m^2}} f(k), 
%p = \int \frac{d^3k}{(2 \pi)^3} \frac{k^2}{3 \sqrt{k^2+ m^2}} f(k). 
\end{equation}
where $a$ is the scale factor from the FLRW metric and $f(k)$ is the distribution function.    
The local momentum $k$ is defined as the scale factor $a(t)$ times the proper momentum.  We use the definitions for stress-energy components according to \cite{9506072}, 
which account for the curvature of FLRW space since the negative pressure of dark energy is a large-scale effect, and we quantize the field below with this in mind as well.  
%is used throughout this work, and we use definitions of stress-energy components 
%according to \cite{9506072}.
%\cite{KolbTurner} and \cite{Bernstein}. 

%Perhaps the true fundamental particle theory of dark energy (if there is one) would provide us with the true distribution function for dark energy, but until such a theory is obtained, 
%we will using the method that follows.  
To derive a suitable distribution function for scalar-field dark energy, we find the vacuum expectation value 
of the energy density from the stress-energy tensor in 
Einstein's equation and equate it with the expression for the energy density,
\begin{equation}
\label{density}
\rho = \frac{1}{a^4} \int \frac{d^3k}{(2 \pi)^3} \sqrt{k^2+ a^2 m^2} f(k),
%\rho = \int \frac{d^3k}{(2 \pi)^3} \sqrt{k^2+ m^2} f(k),
\end{equation}
to obtain a distribution function $f_\phi$ for the dark energy scalar field $\phi$.  

Assuming a perfect fluid model for dark energy that may be non-minimally coupled to the metric via the coupling $\xi$, the stress-energy tensor component for the energy density is 
\begin{equation}
\label{T00}
T_{00} = \rho_\phi = \frac{1}{2} \dot{\phi}^2 + \frac{1}{2a^2} (\partial_i \phi)^2+ V(\phi) + 3 \xi H^2 \phi^2+ 3\xi H \frac{\partial}{\partial t} \left(\dot{\phi}^2 \right),
\end{equation}
where the "dot" %$\cdot$ 
denotes a derivative with respect to the $t$ coordinate of the metric.  We use $V(\phi) = \frac{1}{2} m^2 \phi^2$, which is compatible with $w \leq -1/3$ for late cosmological times \cite{1804.02987}.  In order to find the expectation value of $T_{00}$, we must quantize the 
field \cite{ParkerToms, 1807.10361}:
\begin{equation}
\label{quantize}
\phi = \frac{1}{\sqrt{2(2 \pi a)^3}} \int d^3k \left[ A_{\vec{k}} e^{i \vec{k} \cdot \vec{x}} h_{\vec{k}}(t) + A^\dagger_{\vec{k}} e^{-i \vec{k} \cdot \vec{x}} h^*_{\vec{k}}(t) \right], ~~~ u \equiv  \frac{1}{\sqrt{2(2 \pi a)^3}},
\end{equation}
where 
\begin{equation}
\label{raiselower}
[A_{\vec{k}},A^\dagger_{\vec{k'}}] = (2 \pi)^3 \delta(\vec{k} - \vec{k'}), ~~~A_{\vec{k}} \ket{0} = 0, ~~~ \bra{0} A^\dagger_{\vec{k}} = 0.
\end{equation}
Using Eqs. (\ref{quantize}) and (\ref{raiselower}) with Eq. (\ref{T00}), we obtain
\begin{align}
\label{VEVrho}
\langle T_{00}  \rangle = \rho_\phi =  (2 \pi)^3 \int d^3k \bigg[ & \frac{1}{2} (\dot{u}^2 |h_{\vec{k}}|^2 + \dot{u} u (h_{\vec{k}} \dot{h}_{\vec{k}}^* +\dot{h}_{\vec{k}} h_{\vec{k}}^*) + u^2 \dot{h}_{\vec{k}} \dot{h}_{\vec{k}}^*)+ \frac{u^2}{2a^2} k^2 | h_{\vec{k}} |^2 + \frac{1}{2} u^2 m^2 | h_{\vec{k}}|^2 \nonumber \\
& + 3 u^2 \xi H^2 |h_{\vec{k}}|^2 + 6 \xi H (u \dot{u} |h_{\vec{k}}|^2 + \frac{u^2}{2} (h_{\vec{k}} \dot{h}_{\vec{k}}^* +\dot{h}_{\vec{k}} h_{\vec{k}}^*) \bigg]. 
\end{align} 

To render this expectation value finite, we regularize and renormalize it via adiabatic subtraction.  Minimizing the action in Eq. (\ref{action}) with respect to $\phi$ results in 
the equation of motion 
\begin{equation}
\label{eom}
\ddot{\phi} + 3 H \dot{\phi} + \frac{dV}{d\phi} + \xi R \phi = 0,
\end{equation}
which assumes that the DM field $\psi$ and the DE field $\phi$ are not dependent on each other since the equation is equivalent to the second part of Eq. (\ref{interactionQ}) with $Q=0$, and this is approximately correct since we the cross section between 
DM and DE is small when DE has very small mass, which is what we will come to shortly.  And for any direct coupling between DE and DM in the Lagrangian, one can   
treat the interaction term in the Lagrangian as perturbative so that Eq. (\ref{eom}) is at least still valid at 0th order.  Substituting Eq. (\ref{quantize}) into Eq. (\ref{eom}), one gets a perturbative solution in adiabatic orders (i.e., orders of derivative of the metric, where a new derivative introduces an extra factor of time coordinate $T$ in the denominator) \cite{ParkerToms}:  
\begin{equation}
\label{h}
h_{\vec{k}} = W^{-1/2} e^{-i \int^t W dt'} + O(T^{-2(n+1)}), 
\end{equation}
where $W$ has non-zero terms for even adiabatic order only, and $W$ in the equation is given to $2n$th adiabatic order.  $W$ can be found iteratively order by order from the relation 
\begin{equation}
\label{recursive}
W^2 = k^2/a^2+m^2 + (6\xi - 3/4)(\dot{a}/a)^2 + (6 \xi-3/2)\ddot{a}/a + W^{1/2} \frac{d^2}{dt^2} W^{-1/2}.
\end{equation}
Using Eq. (\ref{h}) in Eq. (\ref{VEVrho}) and equating Eq. (\ref{VEVrho}) with Eq. (\ref{density}) (a similar process to what is done in \cite{1804.07471}), we find that the distribution function for the DE field $\phi$ is
\begin{align}
\label{f}
f_\phi = & \frac{(2 \pi)^6 a^4}{\sqrt{k^2+a^2 m^2}} \bigg[ \frac{1}{2} \bigg( \dot{u}^2 W^{-1} - \dot{u} u W^{-2} \dot{W} + u^2 (\frac{1}{4} W^{-3} \dot{W}^2 + W) \bigg) + \frac{u^2 k^2}{2a^2} W^{-1} + \frac{1}{2} u^2 m^2 W^{-1} \nonumber \\
& + 3 u^2 \xi H^2 W^{-1} + 6 \xi H (u \dot{u} W^{-1} - \frac{u^2}{2} W^{-2} \dot{W}) \bigg].
\end{align}
%where $\omega \equiv \sqrt{k^2/a^2+m^2}$.  

The method of adiabatic subtraction \cite{ParkerToms} says that for a given quantity $A$, the physical, finite expression
 for $\langle A \rangle$ is $\langle A \rangle  = \sum\limits_i^\infty \langle A \rangle_i - \langle A \rangle_{\rm{divergent ~ orders}}$.  For a slowly varying (i.e., adiabatic) FLRW spacetime, each term in adiabatic order contributes less than the previous term in adiabatic order \cite{ParkerToms}.  So for a given quantity $A$, we can 
approximate by truncating the infinite sum to some sufficiently high adiabatic order.  It turns out that our calculation involving the Boltzmann equation for the interaction term $Q$ will 
result in divergent terms for 6th adiabatic order and lower;  we truncate our expression to 8th order.  We can express each of the terms in Eq. (\ref{f}) as a sum of adiabatic orders, accordingly\footnote{Expressions for these quantities up to 8th order and other important expressions pertinent to this work are contained within a {\it Mathematica} notebook posted on Google Drive at  \text{https://drive.google.com/open?id=1Z-V6HgfN\_c-yEP-MtHpV8\_qtqzIT6ZCQ }.}:
\begin{align}
\label{W_orders}
& W =  \omega^{(0)} + \omega^{(1)} + \omega^{(2)} +  \dots, \nonumber \\
 & W^{-1} = (W^{-1})^{(0)} + (W^{-1})^{(1)} + (W^{-1})^{(2)} + \dots, \nonumber \\
& W^{-2} \dot{W} = (W^{-2} \dot{W})^{(0)} +(W^{-2} \dot{W})^{(1)} +(W^{-2} \dot{W})^{(2)} +\dots, \nonumber \\
& W^{-3} \dot{W}^2 = (W^{-3} \dot{W}^2)^{(0)}+(W^{-3} \dot{W}^2)^{(1)}+(W^{-3} \dot{W}^2)^{(2)}+ \dots ~. 
\end{align}

Using the distribution function in Eqs. (\ref{pressure}) and (\ref{density}), the observational constraints for the present time $t_0$, $\rho_\phi(t_0) = \rho_{DE0}$ and $\frac{p_\phi}{\rho_\phi} \approx -1$, 
can specify $m$ and $\xi$.  Using %$\rho_{DE0} = 4.12 \times 10^{-9}$ Mpc$^{-4}$ 
$\rho_{DE0} = 1.6077 \times 10^{-10}$ eV$^{4}$ from best-fit values from Planck \cite{Planck} and choosing a constant $w_\phi = -0.9$ and the corresponding $a(t)$ for the epoch of DE domination, we obtain 
$m = 1.57 \times 10^{-54}$ eV and $\xi = 0.176$.  
%If we choose a phantom value of $w_\phi = -1.2$, a value that is still observationally plausible \cite{1804.02987}, we obtain $m = 7.62 \times 10^{-54}$ eV and $\xi = -0.194 $.  
A more precise approach, especially in light of the interaction between DE and DM, would involve a global fit for $m$ and $\xi$ along with all other parameters which may or may not assume a constant $w_\phi$ , but since we expect the 
exchange between DE and DM to be small (and since doing a rigorous numerical fit to data is not the purpose of this work), we use these values in our final analysis.  

\bigskip
\begin{center}
{\bf Calculating the Interaction Kernel}
\end{center}

In this section, we write the Boltzmann equation in terms of the continuity equation Eq. (\ref{interactionQ}) in order to identify the interaction kernel $Q$ in terms of the collision term from Boltzmann's equation.  We then calculate the collision term for a tree-level two-to-two conversion via the Yukawa-type coupling we assumed in Eq. (\ref{action}) in terms of the effective distribution function for dark energy we calculated in the previous section.   

The Liouville operator in FLRW space applied to the DM distribution function $f_\psi$ is \cite{Bernstein}
\begin{equation}
\label{Liouville}
\frac{\partial f_\psi}{\partial t} - \frac{\dot{a}}{a} k_i \frac{\partial f_\psi}{\partial k_i} = C(f_\psi),
\end{equation}
where $C(f_\psi)$ is the collision term, and $f_\psi$ is given by the Bose-Einstein distribution function since DM is assumed to be a scalar field:
\begin{equation}
\label{BEdistribution}
f_\psi(k) = \frac{1}{e^{\sqrt{k^2+ M^2}/T}-1}.  
%f_\psi(k) = \frac{1}{e^{\sqrt{k^2+ a^2 M^2}/(aT)}-1}.  
\end{equation}
Technically, the use of the Bose-Einstein distribution is only valid assuming there is no interaction, and we could be more precise by using cosmological perturbation theory to get a first-order correction to the distribution function for dark matter.  However, the collision term we consider will end up being very small, so using the Bose-Einstein distribution here is valid.  Also, the collision term is for local interactions, so we use the local definitions of the stress-energy components and the Bose-Einstein distribution function, according 
to \cite{Bernstein, KolbTurner}.  %This is appropriate also because we will consider a graviton-mediated interaction due to a perturbation around a Minkowski background metric.  
We will 
still utilize our distribution function for dark energy derived using the large-scale definitions of the stress-energy components out of necessity.  In theory, one could consider the 
calculation of a general collision term from an $S$-matrix in curved space \cite{1908.06717} on large scales, but this is not necessary for our purposes here.  
 %Because we do not have an explicit measurement of the present-day DM temperature, we will estimate the DM temperature to be $T \sim 1K$.  

Applying $\int \frac{d^3 k}{(2\pi)^3 }$ to both sides of Eq. (\ref{Liouville}), we obtain
\begin{equation}
\int \frac{d^3 k}{(2\pi)^3 } \dot{f_\psi} - \int \frac{d^3 k}{(2\pi)^3 } \frac{\dot{a}}{a} k \frac{\partial f_\psi}{\partial k} =  \int \frac{d^3 k}{(2\pi)^3 } C(f_\psi),
\end{equation}
where we have used $\frac{d}{dk} = \frac{k^i}{k} \frac{d}{dk_i}$ where $k_i k^i \equiv k^2$.  Using the number density expression, 
\begin{equation}
n_\psi =   \int \frac{d^3k}{(2 \pi)^3 }  f_\psi,
\end{equation}
and integrating the second term on the lefthand side by parts (the boundary term goes to $0$ since $f_\psi \rightarrow 0$ at infinity), we get
the typical form of the Boltzmann equation:
\begin{equation}
\label{Boltzmann}
\dot{n}_\psi + 3 \frac{\dot{a}}{a} n_\psi = \int \frac{d^3 k}{(2\pi)^3 } C(f_\psi) \equiv  B,
\end{equation}
where we have defined $B$ for brevity's sake.  We will consider the interaction $\psi_1 \psi_2 \leftrightarrow \phi_3 \phi_4$ (assuming in Eq. (\ref{action}) that $V(\psi) = \frac{1}{2} M^2 \psi^2 +$ any
higher-order terms which we will ignore).  Including the interaction's symmetry factor of $1/2$, $B$ is given by \cite{Bernstein, KolbTurner}
\begin{align}
\label{B}
B =  &  \int \frac{d^3k_1}{(2 \pi )^3 2\sqrt{k_1^2+M^2}} \frac{d^3k_2}{(2 \pi )^3 2\sqrt{k_2^2+M^2}} \frac{d^3k_3}{(2 \pi )^3 2\sqrt{k_3^2+m^2}} \frac{d^3k_4}{(2 \pi )^3 2\sqrt{k_4^2+m^2}} \times \nonumber \\
& (2\pi)^4 \delta^4(k_1+k_2-k_3-k_4) \left[f_\psi(k_1) f_\psi(k_2) - f_\phi(k_3) f_\phi(k_4) \right] |\mathcal{M}|^2,
\end{align}
which assumes the symmetry of $\mathcal{M}_{12\rightarrow 34} = \mathcal{M}_{34\rightarrow 12}$ and neglects the terms that have three factors of $f$ since these terms are 
comparatively small.  

Since $\rho = n \langle E \rangle$, we can rewrite Eq. (\ref{Boltzmann}) as 
\begin{equation}
\label{rhoBoltzmann}
\dot{\rho}_\psi + 3 H \rho_\psi =  B \langle E_\psi \rangle  + n_\psi \frac{\partial \langle E_\psi \rangle}{\partial t},
\end{equation}
where
\begin{equation}
\label{E}
\langle E_\psi \rangle = \frac{ \int \frac{d^3k}{(2 \pi)^3 } \sqrt{k^2+  M^2} f_\psi(k) }{  \int \frac{d^3k}{(2 \pi)^3 }  f_\psi(k)}
\end{equation}
and 
\begin{equation}
\rho_\psi = \int \frac{d^3k}{(2 \pi)^3 } \sqrt{k^2+  M^2} f_\psi(k).
\end{equation}
Since $\dot{\rho}_\psi + 3 H( \rho_\psi + p_\psi) =Q$ from Eq. (\ref{interactionQ}) and $p_\psi \approx 0$ since we are dealing with cold dark matter, we see from Eq. (\ref{rhoBoltzmann}) that 
\begin{equation}
Q =  B \langle E_\psi \rangle  + n_\psi \frac{\partial \langle E_\psi \rangle}{\partial t} .
\end{equation}
%in order to use a(t) as if the continuity equation were 0 for the individual components, we would really need to use cosmological perturbation theory and have the first-order part 
%of the continuity equation be non-zero.  But what we do is good enough. 
Also, for cold (i.e., non-relativistic) dark matter, $\langle E_\psi \rangle = M$ 
%and $\rho_\psi = M n_\psi$ 
\cite{KolbTurner}, so we can express $Q$ as 
\begin{equation}
\label{Q}
Q = M B. 
\end{equation}
Alternatively, we can multiply both sides of Eq. (\ref{Liouville}) by $\sqrt{k^2 + M^2}$, apply $\int \frac{d^3 k}{(2\pi)^3 }$ to both sides, and integrate by parts the second term on the 
lefthand side, and we obtain
\begin{equation}
\dot{\rho}_\psi + 3 H (\rho_\psi + p_\psi) = \int \frac{d^3k}{(2 \pi)^3} C(f_\psi) \sqrt{k^2+M^2}.
\end{equation}
From this, we have a general expression for the 2-to-2 interaction $Q$ between any two perfect fluids.  However, Eq. (\ref{Q}) is more convenient to use in our case, and we use it in what follows.

The Lorentz invariant scattering amplitude for tree-level Yukawa-type interaction between two scalar fields that appears in $B$ is 
\begin{equation}
|\mathcal{M}|^2=g^4 \left(\frac{1}{M^2-t}+\frac{1}{M^2-u}\right)^2.
\end{equation}
where $s = -\eta_{\mu \nu} (k_1+k_2)^\mu(k_1+k_2)^\nu$, $t= - \eta_{\mu \nu} (k_1 - k_3)^\mu(k_1 - k_3)^\nu$, and $u = - \eta_{\mu \nu} (k_1 - k_4)^\mu(k_1 - k_4)^\nu$.   

Assuming the DM mass is bigger than the DE mass ($M>m$), and working in the center-of-mass frame, we can reduce the $f_\phi(k_3) f_\phi(k_4)$ term of $B$ in Eq. (\ref{B}) to 
\begin{equation}
\label{Breduce1}
f_\phi(k_3) f_\phi(k_4) ~{\rm term} =  \frac{1}{8 (2 \pi)^5} \int_{\sqrt{M^2-m^2}}^\infty dk_3 dk_4 k_4^3 \frac{k_4}{k_3^2+m^2} \sqrt{1-\frac{M^2}{k_3^2+m^2}} f_\phi(k_3) f_\phi(k_4) \mathcal{A}(k_3),
\end{equation}
where the bounds of the integral apply for both $k_3$ and $k_4$, $\mathcal{A}(k) \equiv  \int_{-1}^{1} d(\cos \theta) | \mathcal{M}|^2$, $k_3 \equiv |\vec{k_3}|$, and $k_4 \equiv |\vec{k_4}|$.
Similarly, the $f_\phi(k_1) f_\phi(k_2)$ term of $B$ reduces to
\begin{equation}
\label{Breduce2}
f_\psi(k_1) f_\psi(k_2) ~{\rm term} = \frac{1}{8 (2 \pi)^5} \int_0^\infty dk_1 dk_2 k_2^3 \frac{k_2}{k_1^2+M^2} \sqrt{1-\frac{m^2}{k_1^2+M^2}} f_\psi(k_1) f_\psi(k_2) \mathcal{A}(k_1).
\end{equation}
$\mathcal{A}(k_3)$ evaluates to
\begin{align}
\mathcal{A}(k_3) =  \frac{2 g^4 \left(2 k_3 \left(2 k_3^2+m^2\right) \sqrt{-k_3^2-m^2+M^2}+\left(m^4+4 k_3^2 M^2\right) \tan^{-1} \left[\frac{2 k_3 \sqrt{-k_3^2-m^2+M^2}}{2 k_3^2+m^2}\right]\right)}{k_3 \left(2 k_3^2+m^2\right) \sqrt{-k_3^2-m^2+M^2} \left(m^4+4 k_3^2 M^2\right)},
\end{align}  
and $\mathcal{A}(k_1)$ is
\begin{align}
\mathcal{A}(k_1) = 2 g^4 \left(\frac{2}{4 k_1^2 M^2+\left(m^2-2 M^2\right)^2}+\frac{ \tan^{-1} \left[\frac{2 k_1 \sqrt{-k_1^2+m^2-M^2}}{2 k_1^2-m^2+2 M^2}\right]}{\sqrt{-k_1^2+m^2-M^2} \left(2 k_1^3-k_1 m^2+2 k_1 M^2\right)}\right).
\end{align}

\begin{figure}
\fbox{\includegraphics[scale=1.2]{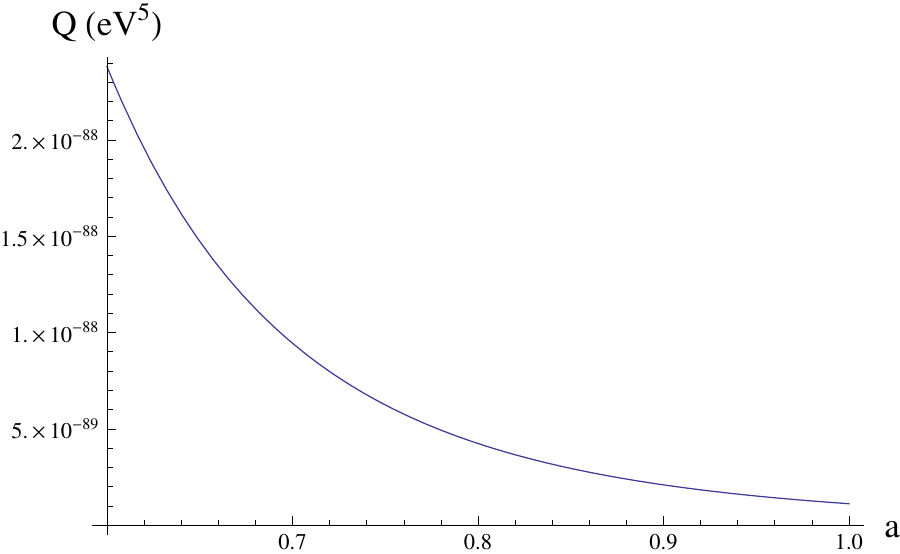}}
\caption{\small{We plot $Q$ (in eV$^{5}$) vs $a$ for $m=1.57 \times 10^{-54}$eV, $M=1$ GeV, $\xi = 0.176$, $g=0.1$eV.  Values of $m$ and $\xi$ chosen by satisfying $\rho_\phi(t_0) = \rho_{DE0}$ and $\frac{p_\phi}{\rho_\phi} = -0.9$ (as constrained by Planck data and discussed above).  The plot is virtually independent of $\xi$ and $m$ (as discussed in the text).  
%$Q$ vs. $a$ assuming DE domination and $w_\phi = -0.9$.  $a=1$ for present-day time.  $m = 1.57 \times 10^{-54}$ eV and $\xi = 0.176$ (as constrained by Planck data and discussed above),  $M = 1$ GeV, and 
%$\chi = 0.1$.  Varying $\chi$ between $1$ and $-1$ hardly affects the plot.
}}
\label{Qa}
\end{figure}

Putting all of this together, we can numerically calculate a finite output for $Q$ as a function of $a(t)$, $m$, $M$, $\xi$, and $g$.  We plot in Fig. (\ref{Qa}) $Q$ as a function of $a$ for the epoch of DE domination, using a typical value for DM mass, $M = 1$ GeV.  The magnitude of $Q$ is very small, indicating a very weak interaction, as 
expected.  
%(For $w_\phi < -1$, $Q$ typically differs in the literature by a minus sign in Eq. (\ref{interactionQ}) compared to the $w_\phi > -1 $ case.  We do not change the sign here 
%so that comparison of the two figures is easier.) 
The $f_\phi(k_3) f_\phi(k_4)$ term contributes essentially $0$ to $Q$, and this is not unexpected given the very low mass value for DE that we use based on the 
observational constraints discussed above.  For virtually any value of $m$,  $-1 \lesssim \xi \lesssim 1$, $1$ GeV $\lesssim M \lesssim 1$ TeV, and $g \lesssim 1$eV, the value of $Q$ is $10^{-100}$ eV$^5 \lesssim Q \lesssim 10^{-80}$ eV$^5$ or smaller\footnote{A phantom value of $w_\phi < -1$ is still observationally plausible even for a normal-sign kinetic term for the scalar field theory \cite{1804.02987}, which is what we use in this paper.  We checked for values of $w<-1$, such as $w=-2$, and the range for $Q$ was the same.}.  
%Also, varying the value of $\chi$ between $1$ and $-1$ hardly affects the plots. 

\iffalse
\begin{figure}
\fbox{\includegraphics[scale=1.2]{Qa_12.pdf}}
\caption{\small{$Q$ vs. $a$ assuming DE domination and $w_\phi = -1.2$.  $a=1$ for present-day time. $m = 7.62 \times 10^{-54}$ eV and $\xi = -0.194 $ (as constrained by Planck data and discussed above), $M = 1$ GeV, and 
$\chi = 0.1$.  Varying $\chi$ between $1$ and $-1$ hardly affects the plot.}}
\label{Qa_12}
\end{figure}
\fi

For comparison, for a typical $Q$ parametrization, $Q = \beta H \rho_{DM}$ from \cite{1812.06854} has a good fit to data for $|Q| \approx - 10^{-45}$ eV$^{5}$ evaluated at the present-day time.  
So one can see that $Q$ for the tree-level 2-to-2 conversion via the Yukawa-type coupling is very small in comparison.  More accurately, we could apply weak equivalence principle constraints to severely limit the magnitude of our coupling $g$ \cite{1605.00996} and thus our magnitude of $Q$, but since the weak equivalence principle is not used to constrain the comparison value of $Q$ from \cite{1812.06854}, we will ignore this for the sake of comparison.
%This is not unexpected since $Q \sim \kappa^{-4}$.  

%Marilena $803.18
%Jason $441.80
%Adam $490.00 (370.58)
%Daniel $176.41

%Greg $435.63
%Ben (Lunch with committee) $143.52
%Total $2490.54

\bigskip
\begin{center}
{\bf Conclusion}
\end{center}

Many seemingly {\it ad hoc} parametrizations are used in the literature to model the interaction between dark energy and dark matter since we do not know the interaction between 
them on a fundamental level.  
%At the very least, dark matter and dark energy should interact gravitationally.  
In this work, we have used adiabatic subtraction to obtain a finite 
analytic expression for the distribution function for dark energy that is compatible with a negative fluid pressure.  Using this distribution function in the Boltzmann equation, we analytically calculate the interaction $Q$ due to the tree-level $2$-to-$2$ 
conversion via our Yukawa-type coupling between dark energy and dark matter.  The interaction is very small compared to typical interaction found in the literature.  

More accurate treatment of our methodology is certainly possible (i.e., using cosmological perturbation theory, doing a global fit to the matter power spectrum and other data), and one can apply this method to other field theories and couplings for DE and DM.  This work serves as a proof of concept that one can treat DE using the formalism of the Boltmann equation with a collision term.
%For future work, we plan to study this coupling as well as other explicit field couplings between dark energy and dark matter and constrain the dark matter mass with a more 
%accurate framework of a global fit to growth of perturbations and other data.      

%\bigskip
%\begin{center}
%{\bf Acknowledgements}
%\end{center}

%KJL is grateful for support from the LaGrange College Summer Research Grant Award.  
%\bigskip
%\begin{center}
%{\bf Appendix}
%\end{center}

\end{document}